\documentclass[ aps,prl,twocolumn,showpacs,superscriptaddress,groupedaddress]{revtex4}

\usepackage{graphics}
\usepackage{makeidx}
\usepackage{epsfig}
\usepackage[caption=false]{subfig}
\usepackage{colortbl}
\usepackage{colordvi}
\usepackage{verbatim}
\usepackage{amsmath, amsthm}
\usepackage{enumerate}
\usepackage{wrapfig}
\usepackage{hyperref}
\usepackage{dcolumn}
\usepackage{bm}
\usepackage{bbm}
\usepackage{amssymb,mathrsfs}
\begin{document}

\title{ Strong Equivalence Principle in Polymer Quantum Mechanics and deformed Heisenberg Algebra}

\author{Nirmalya Kajuri}
\email{nirmalya@imsc.res.in}
 \affiliation{ Department of Physics, Indian Institute of Technology Madras,\\ Chennai 600036, India}

\begin{abstract}
The Strong equivalence Principle (SEP) states that the description of a physical system in a gravitational field is indistinguishable from the description of the same system at rest in an accelerating frame. While this statement holds true in both General Relativity and ordinary Quantum Mechanics, one expects it to fail when quantum gravity corrections are taken into account. In this paper we investigate the possible failure of the SEP in two Quantum Gravity inspired modifications of Quantum Mechanics - Polymer Quantum Mechanics and deformed Heisenberg Algebra. We find that the SEP fails to hold in both these theories. We estimate the deviation from SEP and find in both cases that it is too small to be measured in present day experiments. 
\end{abstract}
\maketitle
\section{Introduction}

The Strong Equivalence Principle (SEP) is the statement that locally, the physics in a uniform gravitational field is identical to the physics in a uniformly accelerated frame. Put differently, SEP says that in the presence of a gravitational field, no matter how strong, locally one can always find a co-ordinate transformation which 'undoes' the gravitational field. The Strong Equivalence Principle is one of the pillars of General Relativity and also holds true in ordinary Quantum Mechanics.

Can it however hold true when Quantum gravity effects are incorporated? One would expect that the answer to be in the negative- quantum gravitational effects should not be undone simply through a co-ordinate transformation. For instance if space is indeed discrete, the particle should 'see' this discrete structure when in the presence of a sufficiently strong gravitational field. Such a particle should not be able to 'un-see' this underlying lattice simply by going to a different frame. Thus one would expect that in the presence of quantum gravity corrections the Strong Equivalence Principle can hold only approximately.  

So it is interesting to ask how the SEP may be modified in the presence of quantum gravitational effects?  As we have argued above, we do expect to see a violation of the SEP. On the other hand, the predicted violation should be small enough to be consistent with the current experimental results, where no violation has been detected \cite{Will:2014kxa, Krauss:1987me, Longo:1987gc, Hohensee:2011wt, Bonder:2013uba, Hohensee:2013cya, Gao:2015lca, Wei:2015hwd}. In this paper, we ask this question for two different modifications of quantum mechanics studied in quantum gravity literature - Polymer Quantum Mechanics and Quantum Mechanics with deformed Heisenberg Algebra. 

Polymer Quantum Mechanics was introduced in \cite{Ashtekar:2002vh}  as a toy model to test features of the quantization technique employed in Loop Quantum Gravity \cite{Thiemann:2007zz, Rovelli:2004tv} in a simple setting (The same quantization had been previously introduced in a different context in \cite{Thirring:1992hu}. It may also be regarded as a physical theory in itself, a theory which incorporates  quantum gravity effects such as spatial discreteness. In this interpretation, various aspects of polymer Quantum Mechanics have been explored and contrasted with the usual Schrodinger Quantum Mechanics. There have been studies of polymer corrections to the dynamics \cite{Husain:2007bj, Kunstatter:2008qx, Demir:2014rfa, Martin-Ruiz:2014rsa, Martin-Ruiz:2015qna} or thermodynamics\cite{ChaconAcosta:2011vv, G.:2013lia, Castellanos:2013ru, Chacon-Acosta:2014zva} in different quantum systems as well comparison with regards to general features such as implementation of symmetries \cite{Date:2012gf}.

It is expected on general grounds (See \cite{Hossenfelder:2012jw} and references therein) that incorporation of quantum gravity effects implies a minimal length or resolution and a modification of the uncertainty principle, leading to Generalized Uncertainty Principles(GUP). The most straightforward and popular quantum mechanical framework for incorporating GUP s is through the modification of the Heisenberg Algebra of position and momentum operators \cite{Kempf:1994su}.Implications of deformed Heisenberg Algebra have been studied extensively in the literature (see \cite{Tawfik:2014zca} and references therein).The gravitational potential in a non relativistic context has been studied in  \cite{Pedram:2012np, Pedram:2013gua, Scardigli:2014qka}. However, while the weak equivalence principle has been investigated in this approach \cite{Ali:2011ap, Tkachuk:2013qa}, there has been no exploration of the SEP till date. 

 We will show that for both polymer Quantum Mechanics and modified Heisenberg Algebra, SEP is violated but the deviation is well within experimental bounds. We will see that SEP is violated in polymer Quantum Mechanics in both the A-polymer and B-polymer representations, as well as in Quantum Mechanics with deformed Heisenberg Algebra However such violations become negligible in scales where quantum gravity effects can be neglected, thus reproducing the result of ordinary Quantum Mechanics. Thus we are obtain a concrete idea about how the SEP, which is one of the cornerstones of General Relativity, is modified in the presence of Quantum gravity effects. This result may have important implications for the unification of Quantum Theory with General Relativity. 

The plan of the paper is as follows. In the next section we present a discussion of the principle of equivalence in non relativistic Quantum Mechanics. In the second section, a derivation of the Strong Equivalence Principle in ordinary quantum mechanics, using extended Galilean transformations. In the third section we first briefly recall Polymer Quantum Mechanics and then proceed to investigate SEP in these theories. The investigation of SEP in QM with modified Heisenberg Algebra is presented in the fourth section. The final section summarizes our results. 
\section{Equivalence Principles in Non Relativistic Quantum Mechanics}
As the principle of equivalence is generally associated with General Relativity, it's appearance in the context of \textit{non-relativistic} quantum mechanics can seem a bit confusing. Another source of confusion may be that there are two different equivalence principles - the strong equivalence principle and the weak equivalence principle - with inequivalent status in quantum mechanics. In this brief section we clarify these issues. 

We start by noting that the statements of both strong and weak equivalence principles are quite independent of relativity. The weak equivalence essentially states that the motion of a particle in a gravitational field can be described without the use of any parameters . In particular, the mass parameter should drop out of the description \cite{Greenberger:2010df}. On the other hand the strong equivalence principle states that the description of motion of a particle in a gravitational system is indistinguishable from the description of a system at rest in an accelerating frame.

Note that both statements are quite independent of relativity and can be made for non relativistic quantum mechanics. Indeed the equivalence principles have been studied in the context of non relativistic quantum mechanics \cite{GREENBERGER1968116, Greenberger:2010df, Viola:1996de, Lammerzahl:1996se, her, davies, accio}. It is found that the weak equivalence principle fails in non relativistic quantum mechanics (see for instance \cite{GREENBERGER1968116}). We will not discuss the weak equivalence principle any further in this paper. 

The general statement of Strong Equivalence Principle given above implies, in the context of non-relativistic quantum mechanics, the following statement: in the Schrodinger equation, if one makes a change of co-ordinates to a frame with acceleration $a$, the only effect will be to add a term equal to $ax$ to the potential. That is, it is indistinguishable from the case where the same quantum mechanical system is being subjected to a constant gravitational force = $a$. This statement can be shown to hold in non relativistic quantum mechanics \cite{Viola:1996de, Greenberger:2010df}. We present the proof in the next section. 

\section{ SEP in Ordinary Quantum Mechanics}

To prove SEP in ordinary Quantum Mechanics we will make use of extended Galileo transformations. This strategy was used in \cite{Greenberger:2010df, Viola:1996de}. However the derivation presented in these references was based on Schrodinger's equation and is not applicable for Polymer Quantum Mechanics, where a differential equation does not appear. Therefore we will present a different derivation based on the action of extended Galileo transformations on operators, which can be extended to the polymer framework. The transformation of operators under Galileo boosts in PQM was first presented in \cite{Chiou:2006rd} - we will extend this treatment to the case of \textit{extended} Galileo transformations. 

We will confine ourselves to extended Galileo transformations between two frames A and A' moving with constant acceleration a with respect to each other. A and A' are assumed to have been coincident as well as moving at the same speed at t=0. For simplicity we will consider only 1 spatial dimension. 

 The action of these extended Galileo transformations on position and momentum operators are 

\begin{align}
\label{position}\hat{B}^\dagger (a,t) \hat{x} \hat{B}(a,t)  = \hat{x} + \frac{1}{2}at^2 \\
\label{moment}\hat{B}^\dagger (a,t) \hat{p} \hat{B}(a,t)  = \hat{p} + mat
\end{align}
Here m is the mass of the particle and t is treated as a parameter.

We may write $B$ as an exponential of its generator: 
\begin{align}
\hat{B}(a,t)= e^{-ia\hat{c}(t)}
\end{align}
Then from \eqref{position} and \eqref{moment} it follows that: 
\begin{align}
\frac{i}{\hbar}[\hat{C}, \hat{x}] = \frac{1}{2} t^2 
\qquad \frac{i}{\hbar}[\hat{C}, \hat{p}] = mt
\end{align}
It follows that 
\begin{align}
\label{glow}\hat{B}(a,t) = e^{\frac{i}{\hbar}[-\frac{t^2}{2}a \hat{p} + ma\hat{x}t]}
\end{align}

Now we will use this definition to prove the equivalence principle. The statement of SEP is the following: The state of a system  accelerated from a frame A to a frame A' at time $t_1$and then evolved till time $t_2$ is the same as the state of a system evolved in frame A from $t_1$ to $t_2$, \textit{but in the presence of an additional constant gravitational field} and then accelerated at time $t_2$ to the frame A'. That is,
\begin{align}
\label{sep}\hat{B}(a,t_2)\hat{U}_{A,a}(t_2, t_1)|\psi \rangle = \hat{U_A'}(t_2, t_1)\hat{B}(a,t_1)|\psi \rangle
\end{align}
where $\hat{U}$ denotes the time evolution operator.$\hat{U_A'}$ is the time evolution operator in the frame A' and $\hat{U_{A,a}}$ denotes the time evolution operator in the frame A in the presence of an additional constant gravitational acceleration a. The various Hamiltonians involved in this case are: 
\begin{align}
\label{h0}\hat{H}_A = \frac{\hat{p}^2}{2m} + V(\hat{x}) \\
\label{h1}\hat{H}_A' =  \frac{\hat{p}^2}{2m} + V(\hat{x - \frac{1}{2}at^2}) \\
\label{h2}\hat{H}_{A,a} =\frac{\hat{p}^2}{2m} + V(\hat{x}) +ma\hat{x}
\end{align}

Taking time derivative with respect to $t_1$ on both sides of \eqref{sep} we have: 
\begin{align}
\label{sept} \hat{B}(a, t_2) \hat{U}_{A,g}  \hat{H}_{A,g} = \hat{H}_A'\hat{U_A'}\hat{B}(a, t_1)\frac{d}{dt_1}\hat{B}(a, t_1)
\end{align}

Now
\begin{align}
\label{dgdt} \frac{d}{dt}\hat{B}(a, t)=\frac{d}{dt}e^{\frac{i}{\hbar}[-\frac{t^2}{2}a \hat{p} + ma\hat{x}t]} = \frac{i}{\hbar}[- ta \hat{p} + ma\hat{x}]\hat{B}(a, t)
\end{align}

Then substituting \eqref{dgdt} in \eqref{sept} and finally using \eqref{sep} we have: 
\begin{align}
\hat{H}_{A,g} =\hat{B}^\dagger (a,t_1) \hat{H}_A' \hat{B}(a,t_1)  - t_1a \hat{p} + ma\hat{x}
\end{align}

But from \eqref{position} and \eqref{moment} we know
\begin{align}
\hat{B}^\dagger (a,t_1) \hat{H}_A' \hat{B}(a,t_1) =  \hat{H}_A'|_{ \hat{x} \rightarrow \hat{x} + \frac{1}{2}at_1^2 , \hat{p} \rightarrow \hat{p} + mat_1}
\end{align}

Then we finally have the following statement of the SEP in Quantum Mechanics: 
\begin{align}
\label{final}\hat{H}_{A,g} = \hat{H}_A'|_{ \hat{x} \rightarrow \hat{x} + \frac{1}{2}at_1^2 , \hat{p} \rightarrow \hat{p} + mat_1} - t_1a \hat{p} + ma\hat{x}
\end{align}

Clearly the Hamiltonians given by \eqref{h1} and \eqref{h2} satisfy \eqref{final}.\footnote{In our derivation the agreement is up to an additive constant, but such a constant has no observable significance.}. Thus we have proven that the Strong Equivalence Principle holds in usual quantum mechanics, at least for all Hamiltonians which have the form \eqref{h0}. Now we turn to the case of SEP in Polymer Quantum Mechanics. 

\section{SEP in Polymer Quantum Mechanics}

\subsection{Polymer Quantum Mechanics}
Let us very briefly recall the basics of polymer representations.

In the Schrodinger as well as Polymer quantizations, the Hilbert Space carries a representation of the Weyl Algebra.
\begin{align}
&W(\zeta_1)W(\zeta_2) = e^{\frac{i}{2} Im\zeta_2\bar{\zeta_2}}W(\zeta_1+\zeta_2) \\
&(W[\zeta])^\star = W[-\zeta]
\end{align}
where $\zeta \in \mathbb{C}$. 

To present these representation we first introduce a length scale $d$ such that 
$$ \zeta = \mu d+ i\frac{\lambda}{d}$$

The Schrodinger representation of Weyl algebra is then given by: 
\begin{align}
&\hat{W}(\mu d) |x\rangle = e^{\frac{i}{\hbar} \mu x}\\
&\hat{W}(i\frac{\lambda}{d})|x\rangle =|x + \lambda\rangle
\end{align}

where the representation is \textit{continuous} in $\mu$ and $\lambda$. This continuity property allows us to define position and momentum operators in the Schrodinger representation.

 Polymer representations are also representations of Weyl algebra with one crucial difference - they are \textit{discontinuous} representations. A representation of Weyl Algebra may be discontinuous in either $\mu$ or $\lambda$. The two possibilities lead to two different representations, known in the literature as a-polymer and b-polymer representations, or simply a and b representations \cite{Corichi:2007tf}. 

 The a-polymer representation is one where the representation of the Weyl algebra is discontinuous in the real part of the argument, $\mu$.
\begin{align}
&\notag\hat{W}(i\frac{\lambda}{d})|p\rangle =e^{i\lambda p }|p\rangle \\
& \implies \hat{p}|p\rangle = {p}|p\rangle 
\end{align}
\begin{align}
\hat{W}(\mu d) |p\rangle = |p +\mu \rangle
\end{align}

Because of the discontinuity in $\mu$, a position operator cannot be defined in the a-polymer representation.

An approximate position operator can be defined by introducing a scale $\mu_0$ .
\begin{align}
\hat{x}_{\mu_0} = \frac{ \hat{W}(\mu_0 d) -   \hat{W}(-\mu_0 d)}{2\mu_0 i}
\end{align}

On the other hand, in the b-polymer representation the discontinuity is in the imaginary part of the argument, $\lambda$ and it is momentum which cannot be well defined.
\begin{align}
&\hat{W}(\mu d) |x\rangle = e^{\frac{i}{\hbar} \mu x}\\
\implies \hat{x}|x\rangle = |x\rangle \\
&\hat{W}(\frac{i\lambda}{d})|x\rangle =|x + \lambda\rangle
\end{align}

Again an approximate momentum operator can be defined by introducing a scale $\lambda_0$.  
\begin{align}
\hat{p}_{\lambda_0} = \frac{\hat{W}(\frac{i\lambda_0}{d}) - \hat{W}(- \frac{i\lambda_0}{d})}{2i\lambda_0}
\end{align}

Thus both the representations introduce a fundamental length. The difference is that the b-representation closely approximates usual quantum mechanics when the corresponding fundamental length $\lambda$  is taken to be very small, while for the a-representation a close approximation  is achieved when the relevant fundamental length $\lambda$  is taken to be very large. 

Thus the b-polymer representation may be thought to be introducing fundamental lattice spacing (or UV cut-off) $\lambda$ analogous to the area quantization in LQG \footnote{ Note however that area is defined as an operator only on the Kinematical Hilbert Space in LQG and therefore it cannot be said that LQG predicts that physical area is quantized}. Correspondingly, the a-polymer representation seems to introduce a maximum length (or IR cut-off) $\mu^{-1}$. This has no analogue in usual LQG, but a maximum bound on the area does appear in q-deformed LQG \cite{Major:1995yz}. 

Interestingly, there is a sense in which two representations are dual to one another \cite{Corichi:2007tf}. This duality bears some resemblance to UV/IR duality of string theory. 

Now we come to extended Galileo transformations in Polymer Quantum Mechanics. In the previous section we saw that the unitary transformation corresponding to an extended Galileo boost was given by:
\begin{align}
\hat{B}(a,t) &= e^{\frac{i}{\hbar}[-\frac{t^2}{2}a \hat{p} + ma\hat{x}t]} \\
&= \hat{W} \left (\frac{matd}{\hbar} -\frac{iat^2}{2d}\right)
\end{align}
As this operator is well defined in Polymer Quantum Mechanics and one may think that we use this as the definition of an extended Galileo boost in the Polymer representations. But note that in \eqref{sept} we differentiated the operator $\hat{B} (a,t)$ with respect to t. But $\hat{W}$ above is \textit{discontinuous} with respect to t, in both polymer representations. This presents a challenge for defining extended Galileo boosts in Polymer Quantum Mechanics. In fact the same issue arises in defining ordinary Galileo boosts in Polymer Quantum Mechanics, though only in the b representation\cite{Chiou:2006rd}. To resolve this, one needs to regularize the extended Galileo boosts appropriately for each representation. In fact the same issue arises for ordinary Galileo boosts, though only for the b representation. An adequate regularization had been given in that case in \cite{Chiou:2006rd}. In this work we will use the same regularization to defined extended Galileo boosts in the b-representation and adopt a similar regularization for the a-representation. 

\subsection{SEP in a-Polymer Representation}

We have stated earlier that we expect the SEP to fail when quantum gravity effects are taken into account. However the way the SEP fails is expected to differ in the a and b representations owing to the different ways a fundamental length scale enters the two theories. In case of the a representation, the fundamental length is a maximal length (alternately a lattice spacing in momentum space). The presence of such an IR cut-off means that it now matters where one chooses the origin to be. The physics given by the a representation would start differing significantly from usual Quantum Mechanics at large distances from the origin. So one would expect that the SEP would be approximately true for a wave function which is localized near the origin and fail for wave functions which are either localized  or spread to points far away (at distances of the order $\mu_0^{-1}$) from the origin. We will see that this expectation is borne out.

As noted in the last section, we will need to define extended Galileo boosts with some regularization. We use the following regularization:
\begin{align}
\hat{B}_{\mu_0}(a,t)= e^{\frac{i}{\hbar}[\hat{x}_{\mu_0}+\widetilde{mat}]}\hat{W} \left (\frac{[mat]d}{\hbar} -\frac{iat^2}{2d}\right)
\end{align}
where $[mat] =\max\left\{n\mu_0\hbar/ m|n\in{\mathbb Z},n\mu_0\hbar/ m\leq mat\right\}$ and the remainder $\widetilde{mat}=mat-[mat]$.

With this regularization we have: 
\begin{align}
\label{reg}\frac{d\hat{B}_{\mu_0}}{dt}(a,t) = \frac{i}{\hbar}[ma\hat{x}_{\mu_0} - at\hat{p}]\hat{B}_{\mu_0}
\end{align}
Which is the same as in \eqref{dgdt}, except with $\hat{x}$ replaced with $\hat{x}_{\mu_0}$, which is appropriate for this representation.

Under this regularized boost, the position and momentum transform as follows:
\begin{align}
\label{pposition}\notag  \hat{B}_{\mu_0}^\dagger (a,t) \hat{x}_{\mu_0} &\hat{B}_{\mu_0}(a,t)  = \\&\frac{e^{\frac{i\mu_0at^2}{2}} \hat{W}(\mu_0 d) -e^{-\frac{i\mu_0at^2}{2}} \hat{W}(-\mu_0 d) }{2\mu_0i} \\
\label{pmoment} \hat{B}_{\mu_0}^\dagger (a,t) \hat{p} &\hat{B}_{\mu_0}(a,t)  = \hat{p} + [mat] + \widetilde{mat}\hat{\alpha}_{\mu_0}
\end{align}

where $$\hat{\alpha}_{\mu_0} = \frac{ \hat{W}(\mu_0 d) +  \hat{W}(-\mu_0 d)}{2} $$

With this regularization the deviation of 

Now let us see the extent to which the Strong Equivalence Principle holds in the a-Polymer representation. Following the same steps as in section II and using \eqref{reg}, \eqref{pposition} and \eqref{pmoment} we have the following statement of SEP in a-Polymer Quantum Mechanics: 
\begin{align}
\notag  &\frac{\hat{p}^2}{2m} + V(\hat{x}_{\mu_0}) + ma\hat{x}_{\mu_0} = \\ \notag &\frac{1}{2m} \left(\hat{p}+ [mat] + \widetilde{mat}\hat{\alpha}_{\mu_0} \right)^2  + ma\hat{x}_{\mu_0} - at\hat{p} \\&+V\left(\frac{e^{\frac{i\mu_0at^2}{2}} \hat{W}(\mu_0 d) -e^{-\frac{i\mu_0at^2}{2}} \hat{W}(-\mu_0 d) }{2\mu_0i}   \right)
\end{align}

Clearly the two sides don't match and SEP does not hold exactly in the a representation of Polymer Quantum Mechanics. Now let us estimate the extent of SEP's failure in this case. 

The right hand side can be re-written as (leaving out a constant term) :
\begin{align}
\notag& \frac{\hat{p}^2}{2m} + 2 \left(\frac{\hat{p}}{2m} + mat\right) \left( 1 - \hat{\alpha}_{\mu_0} \right)\widetilde{mat}+ \left(mat\hat{\alpha}_{\mu_0}\right)^2 \\\notag &+ V \left(\hat{x}_{\mu_0}+ \frac{1}{{2 \mu_0 }}\mathcal{O}(\mu_0^3a^3t^6)\right)+ ma\hat{x}_{\mu_0} \\  \notag
&=  \frac{\hat{p}^2}{2m} + V (\hat{x}_{\mu_0}) +ma\hat{x}_{\mu_0} + \left(\frac{\hbar k}{2m} +mat\right)\hbar \mu_0\mathcal{O}([\mu_0 x_0]^2) \\& \hphantom{x} +\frac{\partial V}{\partial x} \frac{1}{\mu_0}\mathcal{O}(\mu_0^3a^3t^6) 
\end{align}
 where $x_0$ represents the mean position of the particle and $k$ represents the mean momentum.

Thus we see that the two sides are approximately equal when $x_0 \ll \mu_0^{-1}$ and $at^2 \ll \mu_0^{-1}$ - which says that  (1) the particle wavefunction has its mean position near the origin and (2) the wavefunction has only spread distances much smaller than $\mu_0^{-1}$. 

This shows that SEP fails in the a-Polymer representation of Polymer Quantum Mechanics, but as expected it holds to a good approximation at length scales much smaller than the fundamental maximal length given by $\mu_0^{-1}$.
\subsection{SEP in b-Polymer Representation}

In the b representation, quantum gravity effects show up through the length scale $\lambda_0$ which acts as a minimal length. This scale acts as a UV cut-off. Therefore deviation from standard results would be expected at very small distances (of the order of $\lambda_0$). 

In this case we regularize the extended Boost operator as follows:
\begin{align}
 \hat{B}_{\lambda_0}(a,t) = e^{-\frac{i\hat{p}_{\lambda_0}at}{\hbar}}\hat{W}\left( \frac{matd}{\hbar} - \frac{i}{2d}[at^2]\right)
\end{align}

where $[at^2] =\max\left\{n\lambda_0 |n\in{\mathbb Z},n\lambda_0\leq at^2 \right\}$ and the remainder $\widetilde{at^2}=at^2-[at^2]$

Then
\begin{align}
\label{reg2}\frac{d\hat{B}_{\lambda_0}}{dt}(a,t) = \frac{i}{\hbar}[ma\hat{x} - at\hat{p}_{\lambda_0}]\hat{B}_{\lambda_0}
\end{align}

Which is the same as in \eqref{dgdt}, except with $\hat{x}$ replaced with $\hat{x}_{\mu_0}$, which is appropriate for the b representation.

And the transformations are: 
\begin{align}
\label{bposition} \hat{B}_{\lambda_0}^\dagger (a,t) \hat{x} \hat{B}_{\lambda_0}(a,t)  = \hat{x} + \frac{1}{2}[at^2] + \frac{1}{2}\widetilde{at^2}\hat{\beta}_{\lambda_0} 
\end{align}
\begin{align}
\label{bmoment} \notag \hat{B}_{\lambda_0}^\dagger (a,t) \hat{p}_{\lambda_0} &\hat{B}_{\lambda_0}(a,t)  = \\ & \frac{e^{\frac{i\lambda_0mat}{\hbar}} \hat{W}( \frac{i\lambda_0}{d}) -e^{- \frac{i\lambda_0mat}{\hbar}} \hat{W}(- \frac{i\lambda_0}{d})  }{2\lambda_0i}
\end{align}

where $$\hat{\beta}_{\lambda_0} = \frac{ \hat{W}(\frac{i\lambda_0}{d} ) +  \hat{W}(-\frac{i\lambda_0}{d}  )}{2} $$

Now let us see the extent to which the Strong Equivalence Principle holds in the b-Polymer representation. Using \eqref{reg2}, \eqref{bposition} and \eqref{bmoment} we have the following statement of SEP in b-Polymer Quantum Mechanics: 
\begin{align}
\notag &\frac{\hat{p}_{\lambda_0}^2}{2m} + V(\hat{x}) + ma\hat{x} = \\ \notag &\frac{1}{2m} \left(\frac{e^{\frac{i\lambda_0mat}{\hbar}} \hat{W}( \frac{i\lambda_0}{d}) -e^{- \frac{i\lambda_0mat}{\hbar}} \hat{W}(- \frac{i\lambda_0}{d})  }{2\lambda_0i} \right)^2 + \\ &V\left(\hat{x} + \frac{1}{2}[at^2] + \frac{1}{2}\widetilde{at^2}\hat{\beta}_{\lambda_0}  \right) + ma\hat{x}  - at\hat{p}_{\lambda_0}
\end{align}

Once again the two sides of the equation disagree and SEP does not hold. Again let us estimate the scales at which the deviation is negligible. 

Proceeding as before, it is straightforward to check that the RHS can be written as
\begin{align}
\frac{\hat{p}_{\lambda_0}^2}{2m} + V(\hat{x}) + ma\hat{x} +\mathcal{O}([m\lambda_0at]^3) + \lambda_0\frac{\partial V}{\partial x}\mathcal{O}([k\lambda_0]^2)
\end{align}

So the deviations from SEP will be negligible when $m\lambda_0at \ll 1 $ and $k\lambda_0 \ll 1$. Thus the SEP is expected to hold to a good approximation as long as the momentum scale is much less than $\hbar\lambda_0^{-1}$ (alternately at length scales much larger that $\lambda_0$).
\section{Deformed Heisenberg Algebra and the SEP}
\subsection{Quantum Mechanics with Modified Heisenberg Algebra}
In this section we briefly recall the basics of Quantum Mechanics with modified commutation relations. 
In this paper we consider the following deformation of the Heisenberg Algebra \cite{Kempf:1994su}
\begin{equation}\label{com}
[X, P]=i(1+\beta P^2),
\end{equation}
It was shown in \cite{Kempf:1994su} that this leads to a minimal length of resolution: 
 $$\Delta X=\sqrt{\langle X^2 \rangle -\langle X
\rangle^2}\ge\sqrt\beta$$

Thus the relevant length scale is $\sqrt{\beta}$. We would expect quantum gravitational effects to disappear when $\beta$ is small compared to the relevant length scale of the experiment.

The deformed commutation relations \eqref{com} result in the following state dependent generalized uncertainty relations:
\begin{eqnarray}\label{gup}
 \Delta X \Delta P \geq \frac{\hbar}{2}
\left( 1 +\beta (\Delta P)^2 +\gamma \right),
\end{eqnarray}

where $\gamma = \beta \langle p\rangle^2$

The Schrodinger equation is likewise modified. For a Hamiltonian of the form $H=\frac{P^2}{2m} + V(x)$, the Schrodinger equation now becomes a fourth order equation
\begin{eqnarray}\label{H}
-\frac{1}{2m}\frac{\partial^2\psi(x)}{\partial
x^2}+\beta\frac{1}{3m}\frac{\partial^{4}\psi(x)}{\partial
x^{4}} +V(x)\psi(x)=E\psi(x)
\end{eqnarray}

This completes our brief overview of QM with deformed Heisenberg Algebra. In the next section we will investigate the validity of the SEP with the deformed algebra \eqref{com}. One could consider more general deformations, but from our derivation it will be easy to see that similar results would hold for those cases.

\subsection{Investigating SEP in Deformed Heisenberg Algebra}
As in section III we start with the definition of extended Galileo transformations.
\begin{align}
\hat{B}_{\beta}^\dagger (a,t) \hat{x} \hat{B}_{\beta}(a,t)  = \hat{x} + \frac{1}{2}at^2 \\
\hat{B}_{\beta}^\dagger (a,t) \hat{p} \hat{B}_{\beta}(a,t)  = \hat{p} + mat
\end{align}
Once again writing $B_{\beta}$ as 
\begin{align}
\hat{B}_{\beta}(a,t)= e^{-ia\hat{c}_{\beta}(t)}
\end{align}
we have 
 \begin{align}
\frac{i}{\hbar}[\hat{C}_{\beta}, \hat{x}] = \frac{1}{2} t^2 
\qquad \frac{i}{\hbar}[\hat{C}_{\beta}, \hat{p}] = mt 
\end{align}
But as the commutation relations between position and momentum operators has been deformed \eqref{glow} no longer holds -
\begin{align}
\hat{B}_{\beta}(a,t) \neq e^{\frac{i}{\hbar}[-\frac{t^2}{2}a \hat{p} + ma\hat{x}t]}
\end{align}
How do we define the extended Galileo boosts in terms of the basic operators? To do this, we note that if we define an operator $\hat{P}$ such that
\begin{align}
\hat{p} = \hat{P}\left( 1 + \frac{1}{3}\beta\, \hat{P}^2 \right)
\end{align}
Then we will have 
\begin{align}
[\hat{x},\hat{P}]=i\\
[\hat{p},\hat{P}]=0
\end{align}
We then arrive at the following formula for the extended Galilean boost:
\begin{align}
\hat{B}_{\beta}(a,t) = e^{\frac{i}{\hbar}[-\frac{t^2}{2}a \hat{P} + ma\hat{x}t]}
\end{align}
It is then straightforward to follow the steps given in section III. We then arrive at the conclusion that if SEP were to hold in this case the following equation should hold:

\begin{align}
\label{finall}\hat{H}_{A,g} = \hat{H}_A'|_{ \hat{x} \rightarrow \hat{x} + \frac{1}{2}at_1^2 , \hat{p} \rightarrow \hat{p} + mat_1} - t_1a \hat{P} + ma\hat{x}
\end{align}

where as before 
\begin{align}
\hat{H}_A = \frac{\hat{p}^2}{2m} + V(\hat{x}) \\
\hat{H}_A' =  \frac{\hat{p}^2}{2m} + V(\hat{x - \frac{1}{2}at^2}) \\
\hat{H}_{A,a} =\frac{\hat{p}^2}{2m} + V(\hat{x}) +ma\hat{x}
\end{align}

Note that \eqref{finall} differs from \eqref{final} only in that $\hat{p}$ which appeared in \eqref{final} has been replaced by $\hat{P}$ in \eqref{finall}.Thus the left and right hand side of \eqref{finall} don't match - the SEP fails to hold.

It is easy to estimate the failure of SEP. Rhe two sides of \eqref{finall}  differ by a factor of $at_1(\hat{p}-\hat{P}) = \frac{1}{3}at_1\beta \hat{P}^2$. 

This shows that although the SEP fails to be strictly true the deviations can be neglected as long as $at_1\beta$ is small. So, as long as the scale of the experiment is large compared to the deformation scale $\beta$, we don't expect to observe violation of the SEP.

\section{ Summary}
In this paper we have investigated how the Strong Equivalence Principle gets modified in the presence of quantum gravity corrections. We considered two different frameworks -  polymer quantum mechanics and deformed Heisenberg Algebra. In both cases we   found that the Strong Equivalence Principle is violated, but the violations are of the order of appropriate length scales. In case of polymer quantum mechanics, the violation is of the order of the  polymer scale ($\mu_0$ or $\lambda_0$). In case of deformed Heisenberg algebra the violation is of the order of the scale of deformation $\beta$. This suggests that the existing tests of Strong Equivalence Principle should put lower bounds on the values of $\mu_0$, $\lambda_0$ and $\beta$.


\end{document}